\documentclass[10pt,journal,compsoc]{IEEEtran}

\usepackage[T1]{fontenc} 
\usepackage[utf8]{inputenc} 

\usepackage{amsmath,cite,url}

\usepackage{graphicx}
\usepackage{color}
\usepackage{xcolor}
\usepackage{booktabs}

\usepackage{lineno}

\newcommand{\eqnref}[1]{Eq.~\ref{#1}}
\newcommand{\figref}[1]{Fig.~\ref{#1}}

\begin{document}

\title{Music Similarity Learning in Sparse Latin Scenarios}

\author{Gabrielly~Xavier~dos~Santos,
        ~Lucas~Dias~Cler,
        ~and~Olga~Goussevskaia
\IEEEcompsocitemizethanks{\IEEEcompsocthanksitem Computer Science Department, Universidade Federal de Minas Gerais (UFMG), 31270-901 Belo Horizonte, Brazil.\protect\\
E-mails: gabyxsantosufmg@gmail.com, lucasdclerr@gmail.com, olga@dcc.ufmg.br
}
}

\maketitle

\sloppy 


\begin{abstract}
In this work we propose a lightweight,
interpretable approach that estimates item-item music similarity from
publicly available metadata. Our hybrid regression model combines
content-based filtering with the classic item-based collaborative filtering. We
test generalization via a zero-shot evaluation on Latin American music,
a genre for which co-occurrence data is scarce. Accurately predicting
similarity for these unseen, sparse pairs shows that our approach is robust to the regional gap affecting many
existing systems, without the audio signal processing or the computational budget that deep multimodal approaches would require.
\end{abstract}

\begin{IEEEkeywords}
Music similarity, Latin American music, Amazon Reviews Dataset, Data sparsity, Regional gap.
\end{IEEEkeywords}
\IEEEdisplaynontitleabstractindextext
\IEEEpeerreviewmaketitle

\section{Introduction}\label{sec:introduction}

Music recommender systems increasingly rely on deep neural architectures
that process raw audio or proprietary embeddings to model item-item
similarity \cite{oramas2017deep, almahairi2015learning}. This trend has
intensified with the recent adoption of large language models (LLMs) for
music recommendation, which now power multimodal and conversational
recommenders \cite{yun2025user} and are being deployed at commercial
scale \cite{mullakhmetov2026queries}, further raising the computational
and infrastructural barrier to entry into competitive music
recommendation research.

These systems reach strong accuracy but are costly to train and deploy,
depending on licensed audio signals or vendor-specific feature stores
that most researchers cannot access \cite{schedl2018current}, creating
both a computational barrier, since training deep audio encoders
requires substantial processing power, and a data barrier, since raw
audio and proprietary metadata are often gated behind commercial
licenses.

Public music corpora, and the models trained on them, are also skewed
toward Western, English-language catalogs: non-Western genres, including
Latin American music, account for a small share of the data used to
train and evaluate MIR systems \cite{mehta2025music}, and similarity-
and embedding-based recommenders have been shown to disadvantage
artists from underrepresented cultural groups \cite{moradi2025culturalDiversity}.
The MIR community has already developed dedicated, lightweight methods
for robust Latin American music analysis under severe data scarcity
\cite{maia2024selective, silla2008latin}, making it a demanding testbed
for cross-cultural generalization.

We argue that competitive, robust music recommendation does not require
deep audio models or proprietary features, and propose a lightweight,
interpretable approach that estimates item-item music similarity from
publicly available metadata. Our hybrid regression model combines
content-based filtering with the classic item-based CF formulation
\cite{sarwar2001item}, using as ground truth a \emph{regularized cosine
similarity} that applies the ratio of pairwise co-occurrence to
individual review counts as a shrinkage parameter, addressing the
estimation noise raw cosine similarity shows under sparsity
\cite{khawar2019cleaned}. The model learns to approximate this target
from metadata- and review-derived features, following prior work that
mines review text for hidden rating dimensions \cite{mcauley2013hidden,
ni2019justifying}, but without relying on end-to-end neural encoders
\cite{almahairi2015learning}. We evaluate it using RMSE, Spearman
correlation, and NDCG@\{5,10\} \cite{jarvelin2002cumulated}, and further
test generalization via a zero-shot evaluation on Latin American music,
a genre for which co-occurrence data is scarce. Accurately predicting
similarity for these unseen, sparse pairs shows that our hybrid,
non-parametric approach is robust to the regional gap affecting most
existing systems, without the audio signal or computational budget that
deep multimodal approaches require \cite{oramas2017deep}.
\section{Related Work}
\label{sec:related_work}

\noindent\textbf{Item-based CF.}
Item-based collaborative filtering, introduced in \cite{sarwar2001item},
showed that pre-computed similarities between co-rated items offer a
scalable, interpretable alternative to user-based CF. Plain cosine
similarity, however, is unreliable when co-occurrence counts are small, a
common situation in long-tail catalogs such as Latin American music.
\cite{khawar2019cleaned} address this using random matrix theory, showing
that raw cosine similarity overestimates its dominant eigenvalue in noisy,
high-dimensional memory-based CF settings, and propose a spectral
correction.

\noindent\textbf{Mining review text for recommendation.}
Textual reviews are a freely available signal that can complement or
replace sparse rating matrices. \cite{mcauley2013hidden} combine topic
models over review text with latent rating factors on Amazon review data,
the same broad source we build on; their dataset was later expanded in
\cite{ni2019justifying}, which underlies much subsequent review-based
recommendation work. \cite{almahairi2015learning} instead use distributed
neural representations of review text, outperforming bag-of-words topic
models on the same data, exemplifying the accuracy/interpretability
trade-off of neural encoders that our approach avoids.

\noindent\textbf{Multimodal and hybrid music recommendation.}
Hybrid music recommenders combine collaborative signals with content or
text modalities to address cold-start and sparsity. \cite{oramas2017deep}
combine artist biography embeddings with audio-derived track embeddings in
a deep multimodal network, while \cite{oord2013deep} predict CF latent
factors directly from raw audio via convolutional networks; both require
raw audio and non-trivial training infrastructure. \cite{ferwerda2021sequence}
propose a sequence-based, explainable hybrid recommender targeting a
similar accuracy-interpretability trade-off, but evaluate only on
Western-catalog data without a cross-cultural or zero-shot test.
Interpretability and long-tail robustness have also been pursued through
graph-based artist similarity: \cite{gabbolini2021interpretable} propose a
path-based measure over an artist knowledge graph with natural-language
explanations, evaluated with NDCG@10 on small, densely-curated sets, while
\cite{korzeniowski2022artist} combine artist graphs with content features
in a graph neural network targeting long-tail artists, reporting NDCG@200
over much larger candidate pools. Neither addresses cross-cultural or
zero-shot generalization.

\noindent\textbf{Regional gap in MIR.}
Music datasets and the models trained on them are often skewed toward
Western, high-resource catalogs. \cite{schedl2018current} survey open
challenges in music recommender systems, identifying data sparsity and
cultural bias, while \cite{mehta2025music} quantify this imbalance
directly, finding that only a small fraction of hours across widely used
music datasets come from non-Western genres, including Latin American
music. Recent mitigation strategies include socially-motivated
recommendation to counter popularity bias \cite{lacker2024socially} and
embedding-space regularization to offset the disadvantage that
prototype-based matrix factorization imposes on underrepresented cultural
groups \cite{moradi2024advancing}. We address the same problem from an
opposite angle, avoiding learned embeddings and testing generalization
through a zero-shot evaluation on unseen, sparse Latin American item
pairs. Latin American music has already been the focus of dedicated
ISMIR benchmarking and low-resource efforts, including the Latin Music
Database for genre classification \cite{silla2008latin} and
selective-annotation for beat tracking under limited data
\cite{maia2024selective}. Our contribution extends this direction to the
recommendation setting, which, to our knowledge, has not been explicitly
targeted by prior hybrid CF work.
\section{Dataset and Data Preprocessing}

The dataset was derived from the Amazon Product Data collection
\cite{hou2024bridging}, merging the \texttt{CDs\_and\_Vinyl} and
\texttt{Digital\_Music} category subsets to establish the musical domain.

\subsection{Item Metadata and Feature Engineering}
\begin{figure}[h]
 \centerline{\framebox{
 \includegraphics[width=0.95\columnwidth]{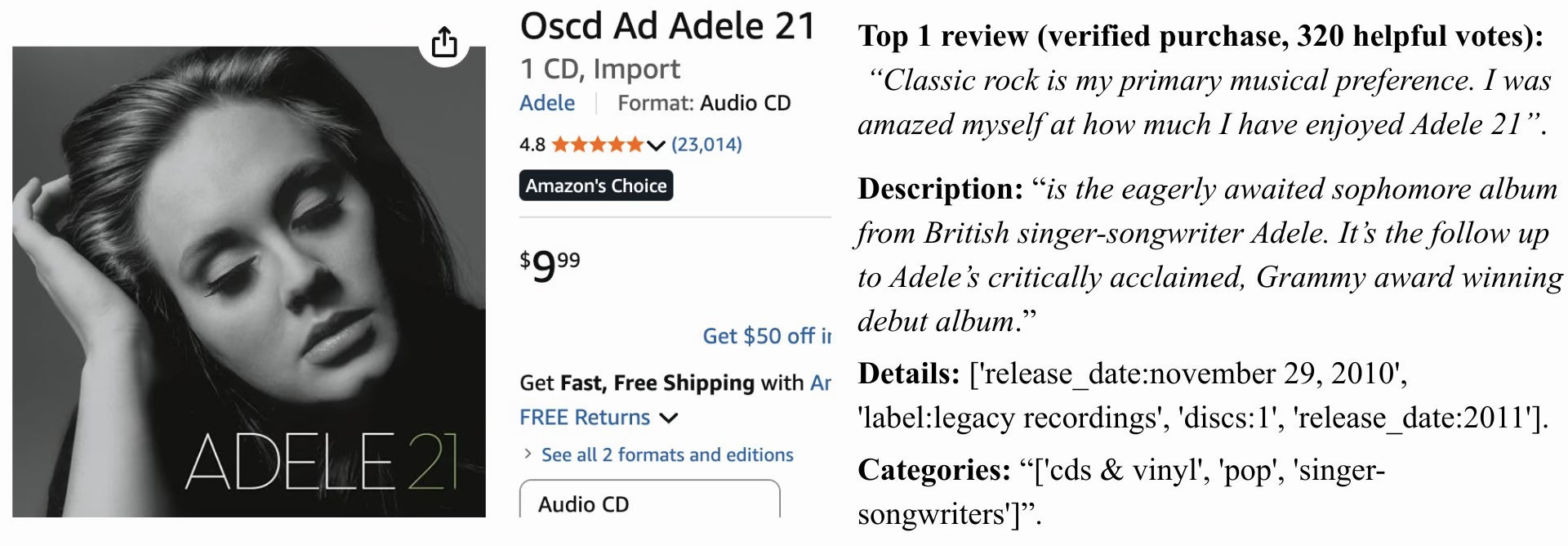}}}
 \caption{Example of a music item and its attributes.}
 \label{fig:item_example}
\end{figure}

The initial metadata comprised 16 attributes per item, narrowed to 6 core
fields: \texttt{title}, \texttt{parent\_asin}, \texttt{main\_category},
\texttt{description}, \texttt{categories}, and \texttt{details}
(Figure \ref{fig:item_example}). To handle sparsity without discarding items,
we engineered binary flags for missing text or structural fields---
description (37.54\%), details (12.40\%), store (9.36\%), and categories
(9.26\% of items)---and lowercased all text fields for consistency.

\subsection{Review Filtering and Regional Splitting}
Reviews were filtered to interactions linked to valid items, then
temporally de-duplicated by \texttt{timestamp}, retaining only the most
recent interaction per user-item pair.

To evaluate our Zero-Shot framework, we isolated the Latin music niche via
case-insensitive matching for the `latin' tag across the
\texttt{categories} array (yielding 13,109 items), and curated the Global
domain excluding these items to keep the two domains mutually exclusive.
Because our continuous target relies on co-occurrence frequencies
(Regularized Cosine), we applied a strict quality filter to the Global
domain, retaining only items with $\ge 10$ reviews to eliminate cold-start
noise, while leaving the Latin domain unfiltered to reflect real-world
sparsity. Table \ref{tab:sparsity_stats} summarizes these characteristics.

\begin{table}[ht]
 \begin{center}
 \small
 \setlength{\tabcolsep}{3pt}
 \begin{tabular}{lccc}
  \toprule
  \textbf{Metric} & \textbf{Raw Global} & \textbf{Filtered ($\ge 10$)} & \textbf{Latin} \\
  \midrule
  Total Reviews & 4,707,193 & 3,239,399 & 42,465 \\
  \hline
  Unique Users & 1,734,849 & 1,385,000 & 28,333 \\
  \hline
  Unique Items & 747,880 & 84,544 & 13,109 \\
  \hline
  Items with 1 Review & 48.12\% & 0.00\% & 55.66\% \\
  \bottomrule
 \end{tabular}
 \end{center}
 \caption{Data volume and sparsity across domains.}
 \label{tab:sparsity_stats}
\end{table}
\section{Similarity Learning: Methodology}
Our methodology develops a scalable, supervised framework for predicting
item similarities that generalizes to sparse regional scenarios without
relying on acoustic or sensitive user data. The pipeline has three
phases: \textbf{Target Formulation}, deriving a continuous ground-truth
metric (Regularized Cosine Similarity) from user co-occurrences;
\textbf{Feature Engineering}, extracting semantic, collaborative, and
structural representations from metadata and reviews; and \textbf{Model
Training and Evaluation}, training an XGBoost regressor and evaluating it
across global and zero-shot regional scenarios using standard ranking
metrics.

\subsection{Target Formulation}

\textbf{Co-occurrence Analysis and Pair Sampling.} Our framework models
item pairs using user co-occurrence frequencies as the foundational
signal. Given the unique item counts (84,544 Global; 13,109 Latin),
exhaustively enumerating all $\binom{n}{2}$ pairs is computationally
prohibitive and overwhelmingly dominated by zero-co-occurrence pairs, so
we applied domain-specific sampling strategies.

For the Global domain, we retained all pairs with $\ge 3$ co-occurrences
as high-confidence relationships (1,006,988 pairs, $X$), and sampled
one-third of this volume ($X/3 = 335{,}662$) from each of the 0-, 1-, and
2-co-occurrence buckets to represent weaker signals without overwhelming
the model, yielding a final training set of 2,013,974 pairs. For the
sparser Latin domain (our zero-shot target), we retained all pairs with
$\ge 1$ co-occurrence (102,635 pairs, $Y$) and sampled an equal number of
zero-co-occurrence pairs, yielding 205,270 pairs for inference.

\begin{figure}[t]
 \centerline{\framebox{
 \includegraphics[width=0.8\columnwidth]{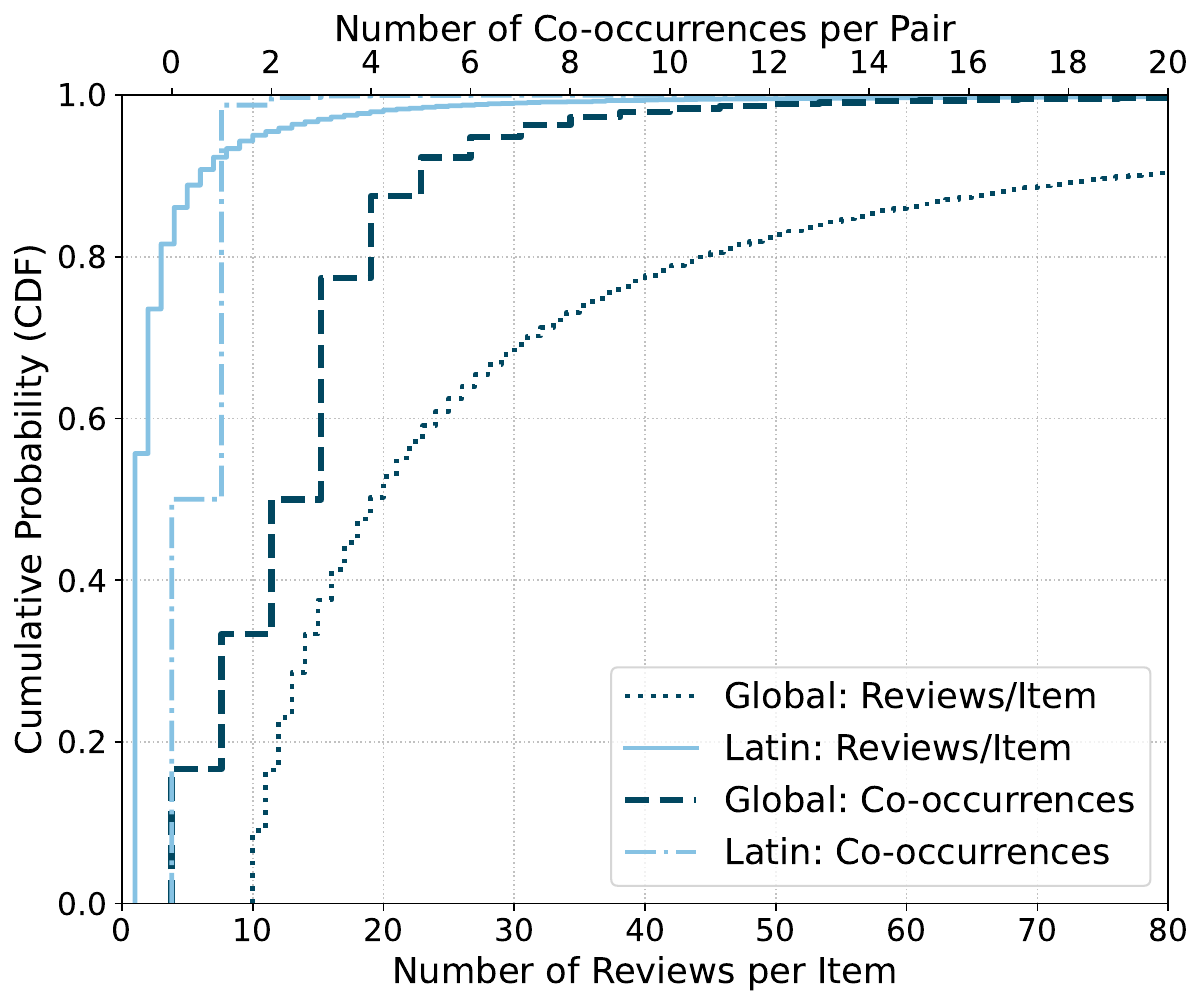}}}
 \caption{CDFs reviews/item and co-occurences/pair.}
 \label{fig:cdf_dupla}
\end{figure}

As illustrated in Figure \ref{fig:cdf_dupla}, the cumulative distributions
capture the effect of the $\ge 10$-review filter on the Global domain
alongside the extreme sparsity and long-tail nature of the Latin niche.

\textbf{Regularized Cosine Similarity.} Let $\vec{v}_a, \vec{v}_b$ be the
binary user-interaction vectors and $N_a, N_b$ the individual interaction
counts of items $a, b$. Standard cosine similarity is biased in sparse
settings, where pairs with very few interactions can receive an
artificial perfect score of $1.0$ simply because they co-occurred; we
mitigate this via a regularized cosine approach \cite{4470228} that adds
a penalty term to the denominator.

Instead of a static constant, we use a dynamic penalty $\alpha_{a,b}$
(\eqnref{eq:alpha}), defined by the proportion of co-occurrences (the dot
product $\vec{v}_a \cdot \vec{v}_b$) relative to each item's individual
frequency:
\begin{equation}\label{eq:alpha}
\alpha_{a,b} = \frac{\vec{v}_a \cdot \vec{v}_b}{N_a} + \frac{\vec{v}_a \cdot \vec{v}_b}{N_b}
\end{equation}

The final regularized similarity score is given in \eqnref{eq:reg_cosine}.
\begin{equation}\label{eq:reg_cosine}
S(a,b) = \frac{\vec{v}_a \cdot \vec{v}_b}{\sqrt{N_a N_b} + \alpha_{a,b}}
\end{equation}

This bounds the maximum penalty at $\alpha = 2$ and scales it per pair,
shrinking false perfect scores caused by severe sparsity and yielding a
more robust ground-truth signal.

\subsection{Feature Engineering}

For each item pair $(a, b)$, we engineer structural, collaborative, and
semantic features to let the regressor approximate the regularized
cosine target.

\noindent\textbf{Structural and Temporal Overlap.} We compute the
Jaccard similarity \cite{leskovec2020mining} between attribute sets of an
item pair to capture shared metadata, and model temporal overlap
\cite{2009korentemporal} via the intersection of each item's active
review years. For attribute sets $\mathcal{A}$ and $\mathcal{B}$, $J(\mathcal{A}, \mathcal{B}) = {|\mathcal{A} \cap \mathcal{B}|}/{|\mathcal{A} \cup \mathcal{B}|}$.


\noindent\textbf{Semantic Embeddings.} We generate dense embeddings
($\mathbf{v}_a, \mathbf{v}_b$) for item descriptions and Top-1 reviews
using a Sentence Transformer (\texttt{all-MiniLM-L6-v2})
\cite{reimers2019sentence}. Reviews are selected via a priority protocol
favoring verified purchases and highest helpfulness scores, falling back
to the most helpful unverified review when needed; missing texts are
left as NaNs to reflect a lack of semantic overlap rather than artificial
proximity. We quantify item proximity via cosine similarity
($\epsilon=10^{-10}$), Euclidean distance, and dot product
\cite{steinbruch1987algebra, manning2008introduction}.

\noindent\textbf{Categorical Matches and Sparsity Flags.} We include a
binary feature for exact artist name matches, and concatenate four
binary missing-value flags (categories, details, store, description) for
each item in the pair (e.g., \texttt{flag\_no\_cat\_A}), enabling the model to adjust predictions based on underlying sparsity.

\subsection{Model Training and Evaluation}

We employ XGBoost (Extreme Gradient Boosting) \cite{chen2016xgboost} as
our regression model, chosen for its ability to capture complex,
non-linear relationships and handle sparse features efficiently without
strict distributional assumptions.

We evaluate three setups: two in-domain configurations
(\textbf{Global$\rightarrow$Global}, \textbf{Latin$\rightarrow$Latin})
and one zero-shot cross-domain scenario
(\textbf{Global$\rightarrow$Latin}) that trains exclusively on Global
pairs and evaluates on Latin pairs. To prevent leakage from items
appearing in both training and testing sets \cite{kaufman2012leakage}, we
performed an item-level 80/20 split prior to pair generation for both
domains, yielding four mutually exclusive item subsets (Global/Latin
$\times$ train/test) from which all in-domain pairs were constructed. For
the zero-shot experiment, domain disjointness naturally prevents leakage,
so we use all Global pairs for training and all Latin pairs for testing.

We assess predictive accuracy and ranking effectiveness using three
complementary metrics: \textbf{RMSE}, the point-wise deviation between
predicted and ground-truth similarity; \textbf{Spearman}, the monotonic
rank correlation between predicted and actual similarity; and
\textbf{NDCG@K} ($K=5,10$), the quality of top-$K$ rankings, which
penalizes errors near the top of the list. Additionally, SHAP (SHapley
Additive exPlanations) \cite{lundberg2017shap} analysis quantifies the
relative impact of different feature groups on the predictions. 
\section{Results}
\noindent\textbf{Machine Learning Performance.} To account for stochastic variance (data splitting, negative sampling,
tree construction), experiments were run across four random seeds
($\{0, 42, 123, 999\}$); the minimal standard deviations reported in
Table~\ref{tab:results} confirm the framework's stability.

\begin{table}[h]
 \begin{center}
 \small
 \setlength{\tabcolsep}{2pt}
 \begin{tabular}{lccc}
  \toprule
  \textbf{Metric} & \textbf{Global $\rightarrow$ Global} & \textbf{Latin $\rightarrow$ Latin} & \textbf{Zero-Shot} \\
  \midrule
  NDCG@10 & 0.731 (0.007) & 0.480 (0.007) & 0.477 (0.007) \\
  \hline
  NDCG@5  & 0.714 (0.008) & 0.470 (0.007) & 0.465 (0.006) \\
  \hline
  Spearman & 0.484 (0.006) & 0.521 (0.014) & 0.400 (0.015) \\
  \hline
  RMSE    & 0.031 (0.000) & 0.118 (0.002) & 0.153 (0.006) \\
  \bottomrule
 \end{tabular}
\end{center}
 \caption{Model performance (mean, standard deviation).}
 \label{tab:results}
\end{table}

The Global$\rightarrow$Global scenario establishes a strong data-rich
baseline (Table~\ref{tab:results}), with high NDCG@10/@5 and a low RMSE
indicating solid point-wise and ranking accuracy. Most notably, the
Zero-Shot NDCG@10 ($0.477$) is statistically competitive with the
in-domain Latin$\rightarrow$Latin baseline ($0.480$), showing that the
model transfers generalized similarity patterns to unseen Latin data
without domain-specific retraining. The counterintuitive increase in
Spearman correlation for Latin$\rightarrow$Latin ($0.521$ vs. $0.484$),
despite degrading RMSE and NDCG, likely reflects differing target
variance across domains: a less continuous similarity spectrum in the
sparse niche makes relative ordering easier to capture than absolute
accuracy.

These results validate our hybrid methodology: integrating SBERT
embeddings with structural metadata and engineered pairwise features
mitigates sparsity, while XGBoost offers a computationally efficient
alternative to deep neural networks with competitive ranking quality at
a fraction of the training cost.

\noindent\textbf{SHAP Analysis.} We partition the feature space into three categories: \textbf{Review
Data} (cosine similarity, Euclidean distance, and dot product between
review embeddings, plus Jaccard temporal overlap); \textbf{Description
Data} (the same vector metrics between description embeddings, plus
description-availability indicators); and \textbf{Other Metadata}
(exact artist matches, Jaccard similarity for \textit{store} and
\textit{categories}, and missing-value indicators for \textit{details},
\textit{store}, and \textit{categories}).

\begin{figure}[t]
 \centerline{\framebox{
 \includegraphics[width=0.85\columnwidth]{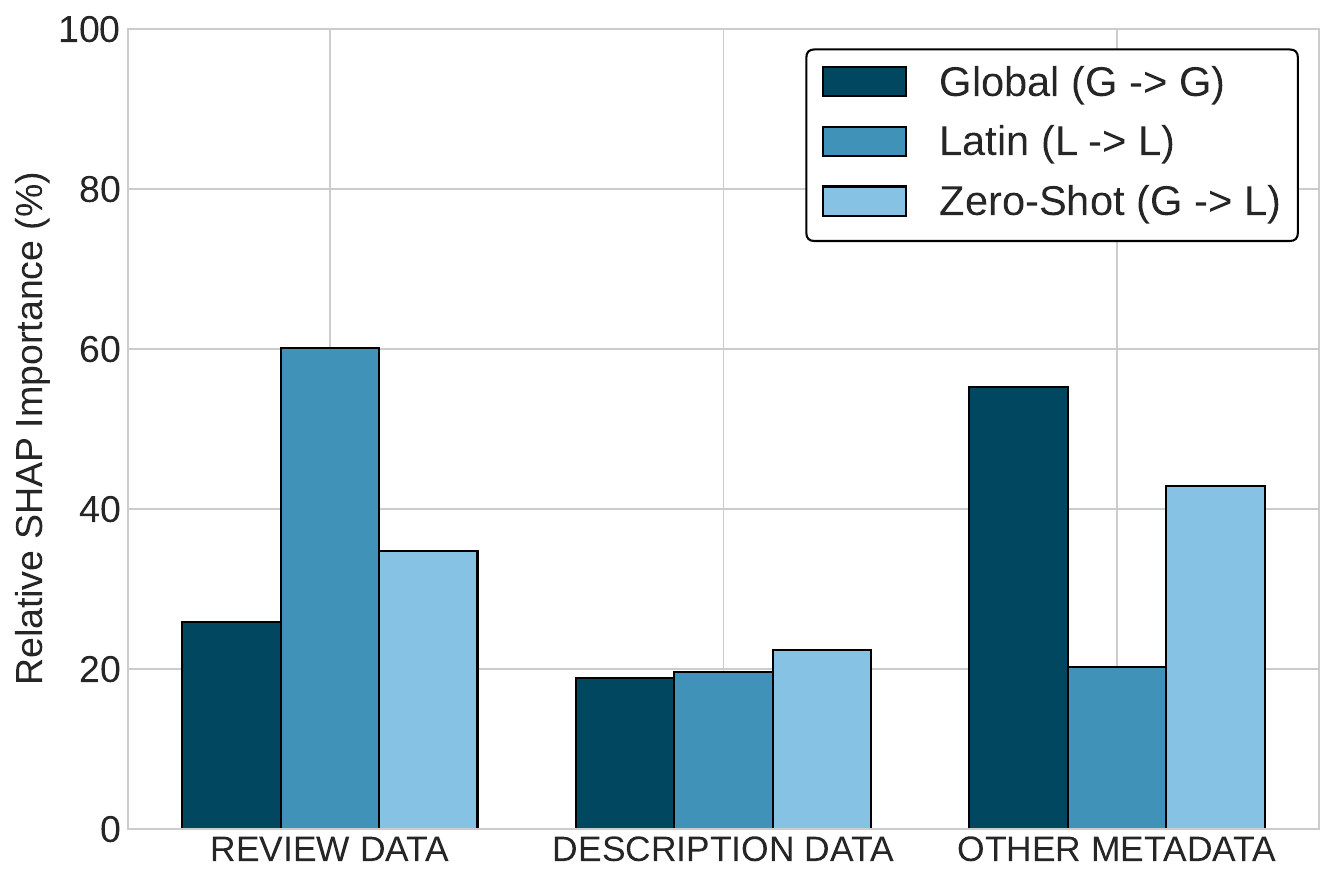}}}
 \caption{SHAP analysis results.}
 \label{fig:shap}
\end{figure}

\figref{fig:shap} reveals a distinct shift in feature reliance between
domains. The Global model relies mainly on `Other Metadata' ($\sim$55\%),
reflecting the reliability and abundance of categorical attributes in
mainstream catalogs, whereas the Latin model shifts drastically toward
`Review Data' ($\sim$60\%), indicating that rare interaction signals
become the most discriminative factor in sparse niches. `Description
Data' remains stable at $\sim$20\% across all setups, highlighting the
domain-agnostic robustness of SBERT-based features. The Zero-Shot model
shows hybrid behavior, retaining a preference for `Other Metadata'
($\sim$43\%) from its metadata-heavy training while redirecting
attention toward `Review Data' ($\sim$35\%) at inference, illustrating
the signal compromises inherent to cross-domain transfer.

\noindent\textbf{Limitations.} Our regional split relies on the \texttt{categories} field---a
structured array of discrete tags (e.g., \texttt{['cds \& vinyl',
'pop']}) rather than free text---flagging an item as Latin American if
any entry contains ``latin''. While more precise than full-text search,
this conflates heterogeneous subgenres (Latin Jazz, Latin Pop, Regional
Mexican) into one label and depends on seller-tagging consistency,
potentially introducing noise or omitting untagged items. Because the
Global domain is filtered to items with $\geq 10$ reviews while the
Latin domain is left unfiltered to preserve realistic sparsity, absolute
metric values across domains are not directly comparable; reported gaps
should be read as evidence of robustness under sparsity rather than
strict performance parity. Finally, our approach relies entirely on
Amazon metadata and reviews, whose signals may differ from
streaming-platform behavioral data (e.g., play counts, skip rates);
validating against such data remains future work.

\section{Conclusion}
We presented a lightweight, interpretable approach for music similarity
estimation using publicly available metadata and review-derived semantic
and temporal features, with a regularized cosine similarity as ground
truth to counteract score inflation from sparse co-occurrence. Across
three scenarios---Global in-domain, Latin in-domain, and Global-to-Latin
zero-shot---our model performs strongly on well-represented data and
remains meaningfully predictive on the sparse, underrepresented Latin
American domain, supporting our claim that robust recommendation does
not require costly audio features or proprietary data. NDCG@K values in
related literature are highly sensitive to candidate-pool size,
ground-truth density, and cutoff $K$, as seen in the gap between
\cite{gabbolini2021interpretable} and \cite{korzeniowski2022artist},
limiting direct comparability across protocols, including ours. Future
work includes validating our methodology on Last.fm-360k
\cite{celma2010music} and the Million Song Dataset
\cite{bertin2011million} under matched protocols, and further
investigating the zero-shot degradation to close the regional gap.
\bibliographystyle{IEEEtran}
\bibliography{ISMIRtemplate}

\end{document}